\documentclass[a4paper]{PoS}

\usepackage[utf8]{inputenc}

\usepackage{amssymb,amsfonts,amsmath,amsthm}
\usepackage{slashed}

\usepackage{dcolumn}

\usepackage{subfigure}

\title{Wave functions of the nucleon and its parity partner from lattice QCD}

\ShortTitle{Wave functions of the nucleon and its parity partner from lattice QCD}

\author{\speaker{Nikolaus Warkentin}$^a$, 
	Vladimir M.~Braun$^a$, 
	Meinulf G\"ockeler$^a$, 
	Thomas Kaltenbrunner$^a$, 
	Andreas Sch\"afer$^{ab}$,
      Gerrit Schierholz$^{ac}$,
      Yoshifumi Nakamura$^c$,
      Dirk Pleiter$^c$,
	Roger Horsley$^d$,
	James M.~Zanotti$^d$,
	Paul E.~L.~Rakow$^e$,
	Hinnerk St\"uben$^f$
\\
        \llap{$^a$}Institut f\"ur Theoretische Physik, Universit\"at Regensburg, \\	93040 Regensburg, Germany\\
	\llap{$^b$}Yukawa Institute for Theoretical Physics, Kyoto University, Japan\\
      \llap{$^c$}Deutsches Elektronen-Synchrotron DESY and John von Neumann Institut f\"ur Computing NIC,\\ 15738 Zeuthen, Germany\\
	\llap{$^d$}School of Physics and Astronomy, University of Edinburgh, \\	Edinburgh EH9~3JZ, UK\\
	\llap{$^e$}Theoretical Physics Division, Department of Mathematical Sciences, University of Liverpool,\\  Liverpool L69~3BX, UK\\
	\llap{$^f$}Konrad-Zuse-Zentrum f\"ur Informationstechnik Berlin,\\ 14195 Berlin, Germany\\
        E-mail: \email{nikolaus.warkentin@physik.uni-regensburg.de}
	}

\author{\centering QCDSF collaboration}

\abstract{
We compute moments of distribution amplitudes using gauge configurations with two flavors of clover fermions from QCDSF/DIK and operators which are optimized with respect to their behavior under the lattice symmetries. The knowledge of these quantities helps in understanding the internal structure of hadrons and in the analysis of (semi-)exclusive processes. We present results for the nucleon distribution amplitude which suggest that the asymmetries (the deviations from the asymptotic form) are smaller than indicated by sum rule calculations. Using the same approach we were also able to calculate the same quantities for the $N^\star(1535)$, the parity partner of the nucleon. These results show a stronger deviation from the asymptotic form.
}

\dedicated{DESY 08-140\\
Edinburgh 2008/33
}

\FullConference{The XXVI International Symposium on Lattice Field Theory\\
		 July 14-19 2008\\
		 Williamsburg, Virginia, USA}

\begin{document}

\section{Introduction}

Distribution amplitudes 
\cite{Chernyak:1977as,Efremov:1979qk,Lepage:1979za} 
describe  the structure of hadrons in terms of valence quark Fock states at 
small transverse separation and are required in the calculation of hard 
(semi)exclusive processes. A simple picture is 
obtained at very large values of the momentum transfer. 
In this limit the process factorizes and, for example, 
the magnetic Sachs form factor of the nucleon $G_M(Q^2)$ can 
be expressed as a convolution of the hard scattering kernel $h(x_i, y_i, Q^2)$ 
and the leading-twist quark distribution amplitude in the nucleon 
$\varphi(x_i, Q^2)$~\cite{Lepage:1979za},
\begin{equation}
      G_M(Q^2)
=f_N^2  \int_0^1 [\mathrm d x] \int_0^1  [\mathrm d y]\;
               \varphi^\star(y_i,Q^2) h(x_i, y_i, Q^2) \varphi(x_i,Q^2),
   \label{eq:formfactor}
   \end{equation}
where $[\mathrm d x]=\mathrm d x_1\mathrm d x_2 \mathrm d x_3\delta(1-\sum_{i=1}^3 x_i)$, and $-Q^2$ is the squared momentum transfer in the hard scattering process. However, in the kinematic region
$1 \, \mathrm{GeV}^2 < Q^2 < 10 \, \mathrm{GeV}^2$, which has attracted a lot
of interest recently due to the JLAB data \cite{Gayou:2001qd,Punjabi:2005wq} for $G_M$, 
the situation is more complicated. Here calculations
are possible, e.g., within the light-cone sum rule approach \cite{Braun:2001tj,Lenz:2003tq}. 
They indicate that higher-twist distribution amplitudes become important while higher Fock 
states do not play a significant role. In any case, the distribution amplitudes are needed as input.

As advocated in the pioneering work \cite{Martinelli:1988xs}, lattice 
QCD is well suited to calculate such non-perturbative quantities.
Our recent calculation \cite{Gockeler:2008xv} of moments of the nucleon 
distribution amplitudes shows that they can be determined reasonably well on the lattice. 
Furthermore, using the same methods we were able to  determine distribution 
amplitudes of the nucleon parity partner $N^\star(1535)$ 
with comparable accuracy. We find that the asymmetry of the leading-twist 
amplitude of the nucleon is smaller than in QCD sum rule calculations, in
agreement with light-cone sum rules \cite{Braun:2006hz} and 
quark models \cite{Bolz:1996sw}, which suggest a less asymmetric form. On the 
other hand, our results for 
$N^\star(1535)$ suggest that the asymmetry for the parity partner of the nucleon 
is considerably enhanced.

\section{Basics}


In this section we work in Minkowski space.  
The leading-twist distribution amplitudes for octet baryons and in particular 
nucleon distribution amplitudes were 
introduced within the classical framework of hard exclusive processes in 
\cite{Chernyak:1977as,Efremov:1979qk,Lepage:1979za} 
 The starting point for baryons is the matrix element of a
trilocal quark operator, which can be written to leading-twist accuracy  
in terms of three invariant functions $V$, $A$ and $T$ \cite{Henriques:1975uh}:
{\small
   \begin{equation*}
      \begin{split}
      \langle 0 \vert &
      \left[
         P \exp \left( i \mathrm g \int_{z_1}^{z_3} A_\mu(\sigma) \mathrm d \sigma^\mu  \right)
         f_\alpha (z_1)
      \right]^a
      \left[
         P  \exp \left( i \mathrm g \int_{z_2}^{z_3} A_\nu(\tau) \mathrm d \tau^\nu  \right)
         g_\beta(z_2)
      \right]^b \,
      h_\gamma^c(z_3)\;
      \vert p \rangle \epsilon^{abc}\\
      =&\frac{1}{4} 
      \left\{
            (p\cdot\gamma C)_{\alpha\beta} (\gamma_5 N)_\gamma f_V V(z_i\cdot p)+ (p\cdot \gamma \gamma_5 C)_{\alpha\beta} N_\gamma f_A A(z_i\cdot p)
            +(i  \sigma_{\mu\nu}p^\nu C)_{\alpha\beta}(\gamma^\mu\gamma_5 N)_\gamma f_T T(z_i\cdot p)
      \right\}.\\&
      \end{split}
   \end{equation*}
}%
Here $a,b,c$ are color
indices, $\vert p\rangle$ denotes a baryon state with momentum $p$ and
$f,\,g,\,h$ are quark fields. We consider these matrix elements for the
space time separation of the quarks on the light cone with $z_i=a_i z$ ($z^2=0$)
and $\sum_i a_i=1$. On the r.h.s. $C$ is the charge conjugation matrix, $f_{(V,A,T)}$ are normalization
constants of the leading-twist distribution amplitudes and $N$ is the baryon
spinor.

In momentum space we have
   \begin{equation}
      V(x_i)\equiv\int V(z_i\cdot p) \prod_{i=1}^{3}  \exp\left(i  x_i (z_i\cdot p)\right) \frac{\mathrm d(z_i\cdot p)}{2\pi},
      \qquad V(x_i)\equiv V(x_1,x_2,x_3),
      \label{eq_daft}
   \end{equation}
and similarly for $A(x_i)$ and $T(x_i)$. The distribution amplitudes $V(x_i)$,\,
$A(x_i)$ and $T(x_i)$ describe the quark distribution inside the baryon as
functions of the longitudinal momentum fractions $x_i$. 
They also depend on the renormalization scale.

The moments of distribution amplitudes
   $
      V^{lmn}=\int_0^1 [\mathrm dx]\; x_1^l x_2^m x_3^n\; V(x_1, x_2, x_3) 
   $ 
 are related to matrix elements of local operators such as
   \begin{align}
      \mathcal V_\tau^{\rho \bar l \bar m \bar n }(0)\equiv&
      \mathcal V_\tau^{\rho (\lambda_1\cdots\lambda_l) (\mu_1\cdots\mu_m) (\nu_1\cdots\nu_n) }(0)=
            \epsilon^{a b c} \;
                                       \left[i^l D^{\lambda_1}\dots D^{\lambda_l}f(0)\right]^a_\alpha 
                                          \;     (C\gamma^\rho)_{\alpha\beta}       \;
       \nonumber\\
              &                        \times
                                       \left[i^m D^{\mu_1}\dots D^{\mu_m} g(0) \right]^b_\beta \;
                                       \left[i^n D^{\nu_1}\dots D^{\nu_n} (\gamma_5 h(0) ) \right]^c_\tau,
      \label{eq:vop}
  \end{align}
by
  $
      P_{LTW}\; \langle 0\vert \mathcal V_\tau^{\rho \bar l \bar m \bar n }(0) \vert p\rangle =
         - f_V V^{l m n} p^\rho p^{\bar l} p^{\bar m} p^{\bar n}  N_\tau (p),
   $ 
with similar relations for $A$ and $T$, see, e.g., \cite{Martinelli:1988xs}.
The multiindex $ \bar l \bar m \bar n $ with $\bar
l\equiv\lambda_1\dots\lambda_l$ and similarly for $\bar m$ and $\bar n$ denotes
the Lorentz structure given by the covariant derivatives $D_{\mu} =\partial_\mu
-i \mathrm g A_\mu$ on the r.h.s. of
eq.~\eqref{eq:vop}. The indices $l,m,n$ (without bars) are the
total number of derivatives acting on the first, second and third quark,
respectively. The index $\rho$ corresponds to the uncontracted Lorentz index
of the gamma matrices in the operators. The leading-twist projection, $P_{LTW}$,
can be achieved, e.g., by symmetrization in Lorentz indices and subtraction of
traces.

Since in the nucleon the two quarks $f$ and $g$ are of 
the same flavor we have additional relations for the moments of the 
distribution amplitudes,
   \begin{equation}
      V^{lmn}=V^{mln},\quad A^{lmn}=-A^{mln},\quad T^{lmn}=T^{mln},
   \end{equation}
which can be restored from the combination
   \begin{equation}
   \phi^{lmn}=\frac{1}{3}(V^{lmn}-A^{lmn}+2T^{lnm})
   \end{equation}
by taking into account the isospin symmetry.
Hence we have only one independent distribution amplitude. In
particular the normalization constants are equal, $f_N=f_V=f_A=f_T$, where $f_N$
is the nucleon wave function normalization constant 
defined by the choice $\phi^{000}=1$. For the parity partner 
of the nucleon $N^\star(1535)$ we have exactly the same relations.
The combination
$\varphi^{l m n}=V^{l m n}-A^{l m n}$, often used in sum rule calculations, can
easily be obtained by the relation
   $
   \varphi^{lmn}=2\phi^{lmn}-\phi^{nml}.
   $ 
 Due to momentum conservation we have additional relations between lower and
higher moments of the distribution amplitudes:
 $  \phi^{lmn}=\phi^{(l+1)mn}+\phi^{l(m+1)n}+\phi^{lm(n+1)}$. 
In particular this implies
\begin{equation}
   1=\phi^{000}=\phi^{100}+\phi^{010}+\phi^{001}=\phi^{200}+\phi^{020}+\phi^{002}+2(\phi^{011}+\phi^{101}+\phi^{110})=\dots
     \label{eq:onesum}
\end{equation}


In the case of next-to-leading twist distribution amplitudes we restrict
ourselves to operators without derivatives. Then only two 
additional constants, $V^0_3$ and $T^0_3$, appear \cite{Braun:2000kw}, which  determine the normalization of the
twist-four distribution amplitudes. 
The combinations $\lambda_1=V_1^0-4V_3^0$ and $\lambda_2=6(V_1^0-4T_3^0)$ are
also used in the literature. They describe the coupling to the nucleon of two
independent nucleon interpolating fields used in the QCD sum rule approach. One
of the operators, $\mathcal L_\tau$, was suggested in \cite{Ioffe:1981kw} and
the other, $\mathcal M_\tau$, in \cite{Chung:1981cc}:
{\small
\begin{align*}
      \mathcal L_\tau (0)&=
      \epsilon^{a b c}  \left[ {u^a}^T(0)  C\gamma^\rho u^b(0) \right ]\times (\gamma_5 \gamma_\rho d^c(0))_\tau,
      &
      \mathcal M_\tau (0)&=
      \epsilon^{a b c}  \left[ {u^a}^T(0)  C\sigma^{\mu\nu} u^b(0) \right ]\times (\gamma_5 \sigma_{\mu\nu} d^c(0))_\tau.
   \end{align*}
}%
Their matrix elements are given by
   \begin{align}
      \langle 0 \vert  \mathcal L_\tau (0) \vert p\rangle &=\lambda_1 m_N N_\tau,&
      \langle 0 \vert  \mathcal M_\tau (0) \vert p\rangle &=\lambda_2 m_N N_\tau.
   \end{align}

\section{Calculation on the Lattice \label{sec:calc}}

It is straightforward to translate the relevant operators to Euclidean space.
However, due to the discretization of space-time, the mixing pattern of the
operators on the lattice is more complicated than in the continuum.
It is determined by the transformation behavior of the operators under
the (spinorial) symmetry group of our hypercubic lattice. As operators
belonging to inequivalent irreducible representations cannot mix, we 
derive our operators 
from irreducibly transforming multiplets of
three-quark operators \cite{Gockeler:2007qs, Kaltenbrunner:2008pb} in order to reduce
the amount of mixing to a minimum. These irreducible multiplets 
constitute also the basis for the renormalization of our operators,
which is performed nonperturbatively in an RI-MOM-like scheme.
In this procedure, the mixing with ``total derivatives'' is automatically taken 
into account. The numerical results presented in this work were obtained using  
{QCDSF/DIK} configurations generated with the standard Wilson gauge action and two flavors 
of nonperturbatively improved Wilson fermions (clover fermions). 
The gauge coupling used was $\beta=5.40$, which corresponds to 
$a \approx 0.067 \, \mbox{fm}$ via a Sommer parameter of $r_0 = 0.467 \, \mbox{fm}$
\cite{Khan:2006de,Aubin:2004wf}. The lattice size was $24^3 \times 48$ with pion masses down to $420\mathrm{MeV}$.

In the case of the moments considered in this work we can avoid
the particularly nasty mixing with lower-dimensional operators completely.
Note that the operators $\mathcal V_\tau^{\rho \bar{l} \bar{m} \bar{n}}$,
$\mathcal A_\tau^{\rho \bar{l} \bar{m} \bar{n}}$ and 
$\mathcal T_\tau^{\rho \bar{l} \bar{m} \bar{n}}$ with different 
multi-indices $\rho \bar{l} \bar{m} \bar{n}$ but the same $l m n$ are 
related to the same moments $V^{l m n}$, $A^{l m n}$ and $T^{l m n}$, and
we make use of this fact not only in order to minimize the mixing problems
but also in order to reduce the statistical noise by considering
suitable linear combinations. 

For the operators without 
derivatives, i.e., the matrix elements $\lambda_1$, $\lambda_2$ and 
$f_N$, we have performed a joint fit of all contributing correlators 
to obtain the values at the simulated quark masses. As these are 
larger than the physical masses a chiral extrapolation to the physical 
point is required in the end. To the best of our knowledge there are no
results from chiral perturbation theory to guide this extrapolation. 
Therefore we have adopted a more phenomenological approach aiming at
linear (in $m_\pi^2$) fits to our data.
It turns out that the ratios $f_N / m_N^2$ and $\lambda_i / m_N$ 
are particularly well suited for this purpose.  
In order to estimate the systematic error due to our linear extrapolation, we 
also consider a chiral extrapolation including a term quadratic in $m_\pi^2$ and 
take the difference as the systematic error.
The results in the $\overline{\mathrm{MS}}$ scheme at a scale of $2 \, \mathrm{GeV}$ 
are given in Table~\ref{tab:latvalues}. 
Note that $2 \lambda_1 \approx - \lambda_2$ for nucleon, a
relation that is expected to hold in the nonrelativistic limit due to Fierz identities.
However, for $N^\star(1535)$ we observe a strong deviation from this relation.

For the higher moments one can proceed in the same way and the constraint
\eqref{eq:onesum} is satisfied very well. However, the
statistical errors in this approach are too large to allow an accurate determination
of the (particularly interesting) asymmetries.  We achieved smaller errors
by calculating the ratios 
$
 R^{l m n}= \phi^{l m n}/S_i
$
 \ where
\begin{align}
 S_1 &= \phi^{100}+\phi^{010}+\phi^{001} &&\text{for $l+m+n=1$,} \\
 S_2 &= 2(\phi^{011}+\phi^{101}+\phi^{110})+\phi^{200}+\phi^{020}+\phi^{002} && \text{for $l+m+n=2$.}
\end{align}
These ratios are extrapolated linearly to the physical masses. 
Again, we extrapolate
all results also quadratically in $m_\pi^2$ and take the difference as an estimate
for the systematic error of the chiral extrapolation.
Requiring that the constraint \eqref{eq:onesum} be satisfied for the renormalized 
values we can finally extract the moments from the ratios. These 
results are summarized in Table~\ref{tab:latvalues}. 
In Table~\ref{tab:firesults} we compare the moments $\varphi^{l m n}$ as obtained 
from $\phi^{l m n}$ with some other estimates. From these values we see that 
the asymmetry of the $N^\star(1535)$ distribution amplitude is more pronounced 
compared to the nucleon and is mostly driven by the first moments. However, in 
both cases we have the approximate symmetry $\varphi^{lmn}\approx\varphi^{lnm}$.

\begin{table}[ht]
 \begin{center}  
\renewcommand{\arraystretch}{1.15}
\vspace{2ex}
\scriptsize 
\centering
\begin{tabular}{  |c||  D{.}{.}{3}
                ||  D{.}{.}{15} |  D{.}{.}{15} |}
\hline
                    & \multicolumn{1}{c||}{Asympt.}&  \multicolumn{1}{c|}{LAT $N$} &  \multicolumn{1}{c|}{LAT $N^\star (1535)$}\\ \hline
$ f_N\cdot 10^{3} [\mathrm{GeV}^2]$      &                       & 3.144(61)(29)(54)   & 4.417(114)(215)(2)   \\
$-\lambda_1\cdot 10^{3} [\mathrm{GeV}^2]$&                       & 38.72(70)(43)(106)  & 40.88(110)(778)(57)  \\
$ \lambda_2\cdot 10^{3} [\mathrm{GeV}^2]$&                       & 76.23(139)(84)(207) & 208.9(47)(384)(42)   \\
\hline
$\phi^{100}$                             & \frac13\approx 0.333  & 0.3638(11)(68)(3)   & 0.4007(14)(48)(12)   \\
$\phi^{010}$                             & \frac13\approx 0.333  & 0.3023(10)(42)(5)   & 0.2610(18)(13)(16)   \\
$\phi^{001}$                             & \frac13\approx 0.333  & 0.3339(9)(26)(2)    & 0.3384(11)(35)(4)    \\
\hline
  $\phi^{011}$                       & \frac17\approx 0.143      & 0.0724(18)(82)(70)  & 0.0706(22)(37)(66)   \\
  $\phi^{101}$                       & \frac17\approx 0.143      & 0.1136(17)(32)(21)  & 0.1213(23)(21)(16)   \\
  $\phi^{110}$                       & \frac17\approx 0.143      & 0.0937(16)(3)(38)   & 0.0943(24)(17)(38)   \\
  $\phi^{200}$                       & \frac{2}{21}\approx 0.095 & 0.1629(28)(7)(68)   & 0.1825(35)(4)(56)    \\
  $\phi^{020}$                       & \frac{2}{21}\approx 0.095 & 0.1289(27)(37)(51)  & 0.0962(37)(68)(119)  \\
  $\phi^{002}$                       & \frac{2}{21}\approx 0.095 & 0.1488(32)(77)(73)  & 0.1429(34)(29)(75)   \\

\hline                                                                           
\end{tabular}
\end{center}                                                                     
\caption{                                                                        
Comparison of our lattice results (LAT) as obtained from {QCDSF/DIK} configurations 
at $\beta=5.40$ for the nucleon $(N)$ and $N^\star(1535)$ at $\mu^2=4\,\mathrm{GeV}^2$
\label{tab:latvalues}. The first error is statistical, the second (third) error 
represents the uncertainty due to the chiral extrapolation (renormalization).
The systematic errors should be considered with due caution, see the text for 
their determination.
}
\end{table}

\begin{table}[ht]
 \begin{center}  
\renewcommand{\arraystretch}{1.15}
\vspace{2ex}
\scriptsize 
\centering
\begin{tabular}{  |c||  D{.}{.}{3}  || D{.}{.}{7} |   
                        D{.}{.}{3}  |  D{.}{.}{3} 
		    ||  D{.}{.}{13} ||  D{.}{.}{13} |}
\hline
                    & \multicolumn{1}{c||}{Asympt.}& \multicolumn{1}{c|}{QCD-SR}&  \multicolumn{1}{c|}{BK} & \multicolumn{1}{c||}{BLW} &  \multicolumn{1}{c||}{LAT $N$} &  \multicolumn{1}{c|}{LAT $N^\star (1535)$}\\ \hline
$ f_N\cdot 10^{3} [\mathrm{GeV}^2]$       &    & 5.0(5)    &   6.64    & 5.0(5)& 3.234(63)(86)   & 4.544(117)(223)   \\
$-\lambda_1\cdot 10^{3} [\mathrm{GeV}^2]$ &    &\multicolumn{1}{c|}{$27(9)$}& &\multicolumn{1}{c||}{$27(9)$}&35.57(65)(136)  &37.55(101)(768)  \\
$ \lambda_2\cdot 10^{3} [\mathrm{GeV}^2]$ &    &\multicolumn{1}{c|}{$54(19)$}& &\multicolumn{1}{c||}{$54(19)$}& 70.02(128)(268) 
& 191.9(44)(391)   \\
\hline
  $\varphi^{100}$ & \frac13\approx 0.333      & 0.560(60) &  0.38      & 0.415 & 0.3999(37)(139)  & 0.4765(33)(155)   \\
  $\varphi^{010}$ & \frac13\approx 0.333      & 0.192(12) &  0.31      & 0.285 & 0.2986(11)(52)   & 0.2523(20)(32)  \\
  $\varphi^{001}$ & \frac13\approx 0.333      & 0.229(29) &  0.31      & 0.300 & 0.3015(32)(106)  & 0.2712(41)(136)   \\
\hline
  $\varphi^{200}$ & \frac17\approx 0.143      & 0.350(70) &  0.18      & 0.212 & 0.1816(64)(212)  & 0.2274(89)(307)   \\
  $\varphi^{020}$ & \frac17\approx 0.143      & 0.084(19) &  0.13      & 0.123 & 0.1281(32)(106)  & 0.0915(45)(224)  \\
  $\varphi^{002}$ & \frac17\approx 0.143      & 0.109(19) &  0.13      & 0.132 & 0.1311(113)(382) & 0.1034(160)(584)  \\
  $\varphi^{011}$ & \frac{2}{21}\approx 0.095 &-0.030(30) &  0.08      & 0.053 & 0.0613(89)(319)  & 0.0398(132)(497)  \\
  $\varphi^{101}$ & \frac{2}{21}\approx 0.095 & 0.102(12) &  0.10      & 0.097 & 0.1091(41)(152)  & 0.1281(56)(131)   \\
  $\varphi^{110}$ & \frac{2}{21}\approx 0.095 & 0.090(10) &  0.10      & 0.093 & 0.1092(67)(219)  & 0.1210(101)(304)  \\
\hline                                                                           
\end{tabular}
\end{center}                                                                     
\caption{
\label{tab:firesults}
Comparison of our lattice results (LAT) for the nucleon $N$ and $N^\star(1535)$ as obtained from {QCDSF/DIK} configurations at $\beta=5.40$ using $\phi^{010}$, $\phi^{001}$, $\phi^{110}$, $\phi^{200}$ and $\phi^{020}$ in Table~\protect\ref{tab:latvalues} to selected sum rule results \cite{King:1986wi} (QCDSR) and the phenomenological estimates \cite{Braun:2006hz} (BLW) and \cite{Bolz:1996sw} (BK) at the scale $\mu^2=1\,\mathrm{GeV}^2$.
}
\end{table}

\section{Model for Distribution Amplitudes}

\begin{figure}
 \subfigure[\label{fig:nlat}]{\includegraphics[width=0.47\textwidth,clip]{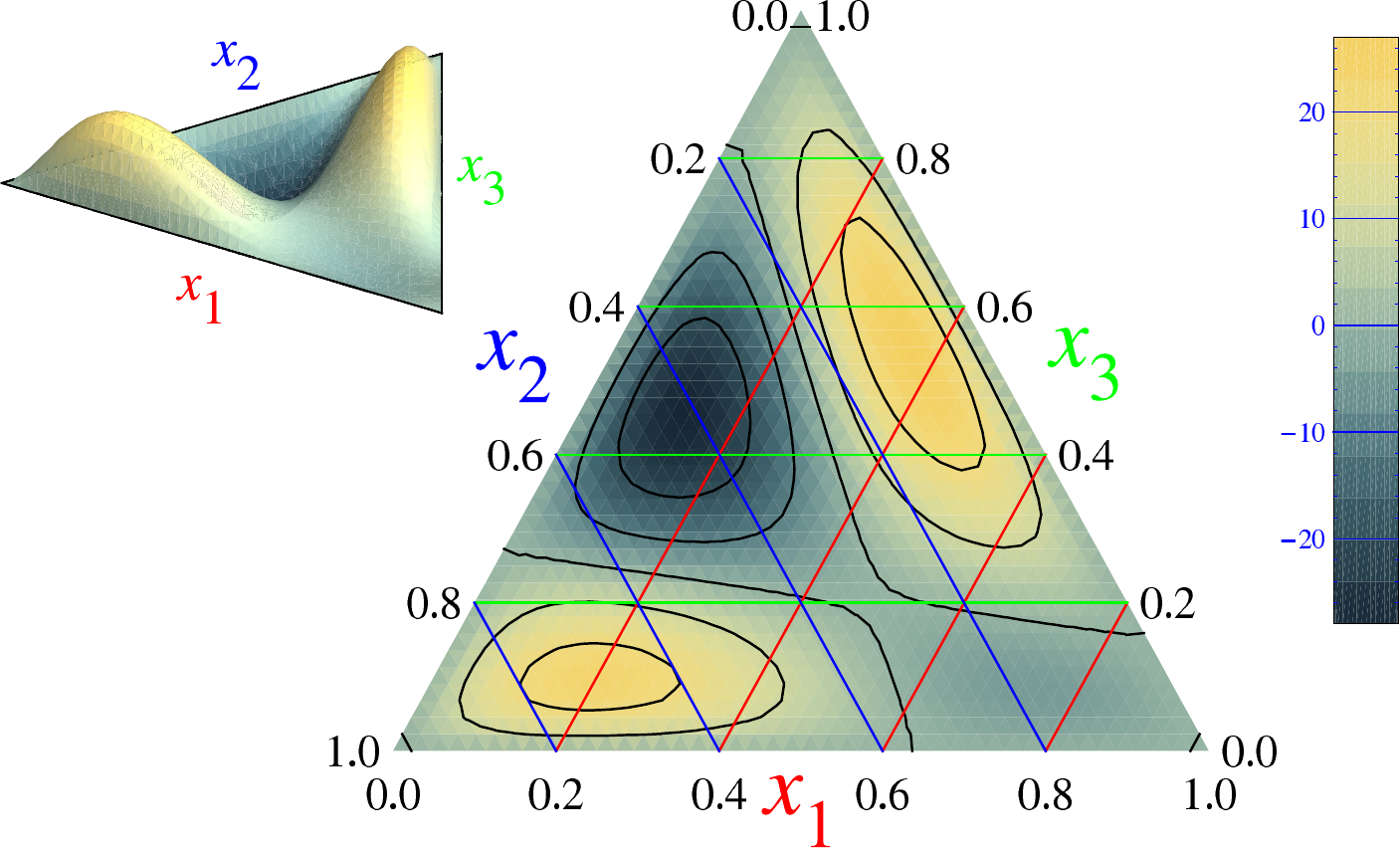}}
 \subfigure[\label{fig:nstar}]{\includegraphics[width=0.47\textwidth,clip]{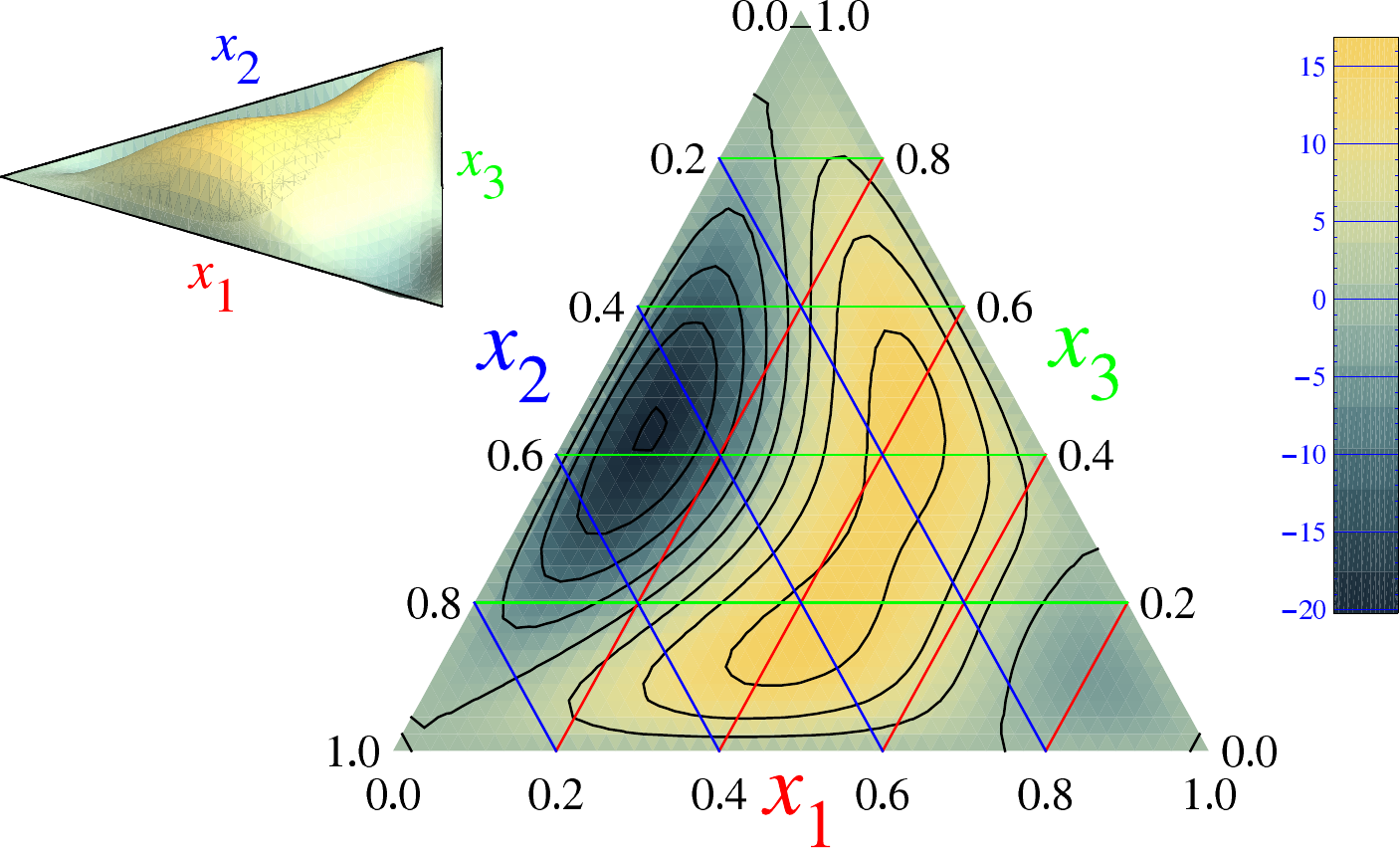}}
\caption{
Barycentric plot of the models of the distribution amplitudes for nucleon (a) and $N^\star(1535)$ (b) at $\mu=1\mathrm{GeV}$ 
using the central values of the lattice results.
The lines of constant $x_1$, $x_2$ and $x_3$ are parallel 
    to the sides of the triangle labelled by $x_2$, $x_3$ and $x_1$, respectively.
}
\end{figure}

Let us now expand the distribution amplitude in terms of orthogonal polynomials 
$P_{nj}$ chosen such that the one-loop mixing matrix is diagonal
\cite{Braun:1999te}:
\begin{equation*}
 \varphi(x_i,\mu)=120 x_1 x_2 x_3 \sum_{n=0}^N \sum_{j=0}^n c_{nj}(\mu_0) P_{nj}(x_i)
\left(\frac{\alpha_s(\mu)}{\alpha_s(\mu_0)}\right)^{\omega_{nj}}.
\end{equation*}
Taking $N=2$ and calculating the coefficients $c_{nj}(\mu_0)$ from an independent subset of
the moments $\phi^{lmn}(\mu_0=2\,\mathrm{GeV})$, we
obtain a model function for the distribution amplitude presented in
Fig.~\ref{fig:nlat}.  
While the (totally symmetric) asymptotic amplitude $120 \, x_1 x_2 x_3$ has a maximum
for $x_1 = x_2 = x_3 = 1/3$, inclusion of the first moments 
(i.e., choosing $N=1$) moves this maximum in the case of the nucleon 
to $x_1 \approx 0.46$,  $x_2 \approx 0.27$,  $x_3 \approx 0.27$
giving the first quark substantially more momentum than the others.
The second moments then turn this single maximum into the two local maxima 
in Fig.~\ref{fig:nlat}. This also happens in the case of $N^\star(1535)$. 
However the influence of the second moments on the form of the DA is reduced 
compared to the nucleon due to the stronger asymmetry in the first moments.
The approximate symmetry in $x_2$ and $x_3$ seen in both cases 
is due to the approximate symmetry 
$\varphi^{lmn}\approx \varphi^{lnm}$ of our results. It is also seen 
in QCD sum rule calculations for the nucleon as well as in several 
models such as BLW and BK.
Since higher-order polynomials have been disregarded,
Figs.~\ref{fig:nlat}~and~\ref{fig:nstar} should be interpreted with
due caution.

\begin{acknowledgments}
We are grateful to A.~Lenz, J.~Bloch and A.~Manashov for helpful
discussions. The numerical calculations have been performed on the Hitachi
SR8000 at LRZ (Munich), apeNEXT and APEmille at NIC/DESY (Zeuthen) and
BlueGene/Ls at NIC/JSC (J\"ulich), EPCC (Edinburgh) and KEK (by the Kanazawa
group as part of the DIK research program) as well as  QCDOC (Regensburg)
 using the Chroma software library \cite{Edwards:2004sx, bagel:2005}. 
This work was supported by DFG 
(Forschergruppe Gitter-Hadronen-Ph\"anomenologie and SFB/TR 55 Hadron Physics from
    Lattice QCD), 
by EU I3HP (contract No. RII3-CT-2004-506078) 
 and by BMBF.
\end{acknowledgments}

\bibliographystyle{utphysm}

\bibliography{nstar}

\end{document}